\documentclass[ 
    reprint,
    superscriptaddress,
    amsmath,amssymb,
    aps,
]{revtex4-1}

\usepackage{graphicx}% Include figure files
\usepackage{dcolumn}% Align table columns on decimal point
\usepackage{bm}% bold math
\usepackage{multirow}
\usepackage{verbatim}
 \usepackage{tikz-feynman,contour}
\usepackage{color}
\usepackage{url}
\usepackage{textcase}

\begin{document}

\title{Flavor and energy inference for the high-energy IceCube neutrinos}

\author{Giacomo D'Amico}
\affiliation{Dipartimento di Fisica, Universit\`a di Roma ``La Sapienza", P.le A. Moro 2, 00185 Roma, Italy}
\affiliation{INFN, Sez.~Roma1, P.le A. Moro 2, 00185 Roma, Italy}

\date{\today}

\begin{abstract}
We present a flavor and energy inference analysis for each high-energy neutrino event observed by the IceCube observatory during six years of data taking. Our goal is to obtain, for the first time, an estimate of the posterior probability distribution for the most relevant properties, such as the neutrino energy and flavor, of the neutrino-nucleon interactions producing shower and track events in the IceCube detector. For each event the main observables in the IceCube detector are the deposited energy and the event topology (showers or tracks) produced by the Cherenkov light  by the transit through a medium of charged particles created in neutrino interactions. 
It is crucial to reconstruct from these observables the properties of the neutrino which generated such event. Here we describe how to achieve this goal using Bayesian inference and Markov chain Monte Carlo methods.
\end{abstract}

\maketitle

\section{\label{sec:introduction} Introduction}

The dawn of the multi-messenger astronomy has been one of the most great achievements in the scientific community in the last decade.  Combining neutrino observations with measurements of cosmic rays, electromagnetic radiation and gravitational wave will be crucial to solve long-standing problems in astrophysics \cite{PhysRevLett.107.251101} and may move forward our current horizons in fundamental physics \cite{Amelino-Camelia:2016ohi}. Nonetheless, the observation of astrophysical neutrinos provides on its own a more deep understanding of neutrino physics.  Neutrino
oscillation phenomena, such as sterile neutrinos \cite{Aartsen:2017bap}, and non-standard interactions \cite{1742-6596-718-6-062011}, are just some topics whose understanding relies in the observation of astrophysical neutrinos.

The largest neutrino telescope  to date is the IceCube Neutrino Observatory at the geographic South Pole, whose first sensors were deployed at the South Pole during the austral summer of 2004-2005 and have been producing data since February 2005 \cite{ACHTERBERG2006155}. After six years of data taking \cite{Aartsen:2017mau}, from early 2010 to early 2016 for a total of 2078 days, 50 neutrino events with deposited energies above 60 TeV have provided the evidence for the existence of an extraterrestrial neutrino flux. Only three events with deposited energy above 1 PeV have been observed, with the 2 PeV event being the most energetic one. The discovery of this flux has motivated a vigorous program of studies to unravel their origin \cite{Aartsen:2017eiu} and their properties \cite{Palomares-Ruiz:2015mka,Aartsen:2016xlq,Aartsen:2013vja}.

IceCube detects neutrinos by observing Cherenkov light produced by
charged particles created in neutrino interactions as they
transit the ice within the detector. At this range of energies, the way neutrinos interact is deep-inelastic scattering with nuclei in the detector material. There are two possible interactions: charged-current (CC) or neutral-current (NC) interactions. In both a cascades of hadrons
is created at the neutrino interaction vertex and for CC interaction this shower is accompanied by an outgoing charged lepton which may itself trigger another overlaid cascades. IceCube events have two basic topologies: tracks and showers. Considering the energy involved for this analysis we assume tracks are made only by $\nu_{\mu}$ CC interactions and by $\nu_{\tau}$ CC interactions in which the tau lepton decays in $\nu_{\tau} \mu \nu_{\mu}$. Showers instead are those events
without visible muon tracks and are formed by particle
showers near the neutrino vertex. While the particle content of showers created by final-state hadrons, electrons, and taus is different, the IceCube detector is currently insensitive to the difference. This means that a shower is produced in $\nu_{e}$ CC interaction, $\nu_{\tau}$ CC interactions (where the produced $\tau$ does not decay in the muonic channel), and  in all-flavor NC interactions. 

In previou works IceCube data have been analyzed and discussed in detail (see Ref. \cite{Aartsen:2017eiu,Chianese:2017jfa,Palomares-Ruiz:2015mka} and references therein) using a maximum-likelihood approach over the whole collection of events. Although useful informations about the energy behavior and the flavor composition has already been explored, it has never been performed an inference analysis of the properties of each single neutrino event. This work differs from previous analyses also for the statistical approach used: having to deal, one by one, with just one single event the frequentist approach is unsuitable and may be misleading. For this reason we prefer the Bayesian approach which we discuss in Sec. \ref{sec:analysis}.

The structure of the paper is as follows. We start in Sec. \ref{sec:Flux} by describing our assumptions on energy and flavor flux. Sec. \ref{sec:scattering} provides a description of the deep-inelastic scattering, the energy-dependent cross section of neutrino-nucleon interactions and the neutrino energy loss. Branching fractions of tau-decay channel, along with the energy distribution of the decay products, are presented in Sec. \ref{sec:tau_decay}. A summary of all parameters used in this analysis and a brief description of the Bayesian method can be found in Sec. \ref{sec:analysis}. Finally in  Sec. \ref{sec:results} we highlight our results and discuss their implications.

\section{\label{sec:Flux} Assumptions on neutrino fluxes}

Working with neutrinos implies the knowledge of its energy and flavor, which are not direct observables. We can only infer this quantities by the deposited energy in the detector and by the event topology. This is possible only if we know \emph{a priori} the expected fluxes of incoming-neutrino energy and flavor. 

We assume an equal spectrum and flux for neutrinos and anti-neutrinos at the Earth. When fitting the current available data,  the assumption that neutrino and anti-neutrino
flavor fractions are the same at the Earth seems reasonable  \cite{Nunokawa:2016pop}. All parameters and their properties, if not otherwise specified, are obtained for the sum of neutrino plus anti-neutrino contributions. For the sake of brevity, here and in the rest of this article, we imply also anti-neutrinos
when we speak of neutrinos and we will refer to both neutrinos and anti-neutrinos as $\nu$. 

The flavor ratio at the Earth $(f_e : f_{\mu} : f_{\tau} )_{\oplus}$  is one of the most studied properties of astrophysical neutrinos. This is due mainly to the fact that the flavor ratio of astrophysical neutrinos is both a probe of the source of high energy cosmic rays and a test of fundamental particle physics. A deviation from the expected flavor ratio at the Earth would be a signal of new physics in the neutrino sector. Consistently with
the  $(1 : 1 : 1 )_{\oplus}$ flavor ratio at Earth commonly expected \cite{Majumdar:2006px} and with the results reported by the IceCube collaboration in Ref. \cite{Aartsen:2015ivb}, a uniformly-distributed prior probability for the neutrino flavor will be used in this analysis. Thus
\begin{equation}
f(\ell) =
\begin{cases}
1/3, & \ell = e \\
1/3, & \ell = \mu \\
1/3, & \ell = \tau ,
\end{cases}
\end{equation}
where $\ell$ is the leptonic flavor and $f(\ell)$ its probability distribution. 

Astrophysical neutrinos from cosmic accelerators are generically expected
to have a hard energy spectrum. Waxman and Bahcall \cite{Waxman:1998yy} predicted a cosmic neutrino flux proportional to $E_{\nu}^{-2}$, as originally predicted by
Fermi. But the spectral index may depends on the source properties and the acceleration mechanism, as pointed out in some recent works (see Ref. \cite{Bell:2013gga} for instance). It is also possible that the neutrino fluxes may be described by more than one component \cite{Chianese:2017jfa,Chianese:2017nwe}. In this analysis we assume a single astrophysical component
parametrized in terms of an unbroken power-law per neutrino flavor described by two parameters, the normalization $\Phi_{\text{astro}}$ at 100 TeV neutrino energy and the spectral index $\gamma$:
\begin{equation}
\Phi(E_{\nu} )_{\nu + \bar{\nu}} =  \Phi_{\text{astro}} \cdot \left( \frac{E_{\nu}}{100 \; \text{TeV}} \right)^{- \gamma}.
\end{equation}
Since in this analysis we are mainly interested in inferring the energy $E_{\nu}$ for each single neutrino, the only parameter that matters for us is the spectral index $\gamma$.  The most recent estimate for the spectral index for high-energy astrophysical neutrino observed by IceCube is given in Ref. \cite{Aartsen:2016xlq}, in which they found that the 6-years data are well described by
an isotropic, unbroken power law flux with a hard spectral index of $\gamma = 2.13 \pm 0.13$. Thus our prior distribution for the spectral index will be a normal distribution with mean value $2.13$ and standard deviation $0.13$:
\begin{equation}
\mathcal{N}(\gamma \, | \, 2.13, 0.13) = \frac{1}{{0.13 \sqrt {2\pi } }}e^{{{ - \left( {\gamma - 2.13} \right)^2 } \mathord{\left/ {\vphantom {{ - \left( {x - \mu } \right)^2 } {0.0338 }}} \right. \kern-\nulldelimiterspace} {0.0338 }}}.
\end{equation}
According to the cut considered by the IceCube collaboration we consider only neutrino events with deposited energy in the range 60 TeV - 3 PeV, which is a cut often used when performing statistical analyses of the astrophysical flux. The minimum deposited energy of 60 TeV is intended to eliminate most of the expected atmospheric muon background events (in our work we neglect the contribution to the neutrino fluxes of atmospheric neutrino), while the limit of 3 PeV deposited energy would discard the Glashow resonance at $E_{\nu} \simeq 6.3$ PeV \cite{PhysRev.118.316}, which should give rise to yet-unobserved events in the few PeV region. From all these considerations we use
\begin{equation}
\frac{(\gamma -1) \, E_{\nu}^{-\gamma}}{(60 \, \text{TeV})^{-\gamma +1} - (3 \, \text{PeV})^{-\gamma +1 } }
\end{equation}
with $E_{\nu} \in$ [ 60 TeV, 3 PeV], as our prior probability distribution for the  neutrino energy $E_{\nu}$.

\section{\label{sec:scattering} Neutrino-Nucleon Deep-inelastic scattering}

\begin{figure}[!h]
\begin{tikzpicture}
\begin{feynman}
\vertex (a) {\(\nu_{\ell}\)};
\vertex [below right=1.5cm of a] (b);
\vertex [above right=1.2cm of b] (c) {\( \ell \)};
\vertex [below =1cm of b, blob] (d) {\contour{white}{$ $}};
\vertex [left=of d] (e) {\(  N \)} ;
\vertex [ right=of d] (f) {\(  X \)};
\diagram* {
(a) -- [fermion, edge label'=$E_{\nu}$] (b),
(b) -- [fermion, edge label'=$(1-y) E_{\nu}$] (c),
(b) -- [boson, edge label'=\(W \)] (d),
(e) -- [fermion, thick] (d),
(d) -- [fermion, thick, edge label'=$y E_{\nu}$] (f),
};
\end{feynman}
\end{tikzpicture}
$\qquad$
\begin{tikzpicture}
\begin{feynman}
\vertex (a) {\(\nu_{\ell}\)};
\vertex [below right=1.5cm of a] (b);
\vertex [above right=1.2cm of b] (c) {\( \nu_{\ell} \)};
\vertex [below =1cm of b, blob] (d) {\contour{white}{$ $}};
\vertex [left=of d] (e) {\(  N \)} ;
\vertex [ right=of d] (f) {\(  X \)};
\diagram* {
(a) -- [fermion, edge label'=$E_{\nu}$] (b),
 (b) -- [fermion, edge label'=$(1-y) E_{\nu}$] (c),
(b) -- [boson, edge label'=\(Z \)] (d),
(e) -- [ fermion, thick] (d),
(d) -- [fermion, thick, edge label'=$y E_{\nu}$] (f),
};
\end{feynman}
\end{tikzpicture}
 \caption{
        Diagrams for charged (left) and neutral (right) current neutrino-nucleon interaction. Time runs from left to right and the flavor index $\ell$ represents $e$, $\mu$, or $\tau$.
    }
    \label{fig:CC_NC_plot}
\end{figure}
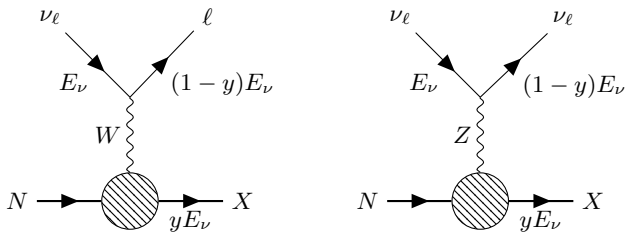

\begin{figure}[!h]
\includegraphics[width=0.9\linewidth,height=0.22\textheight]{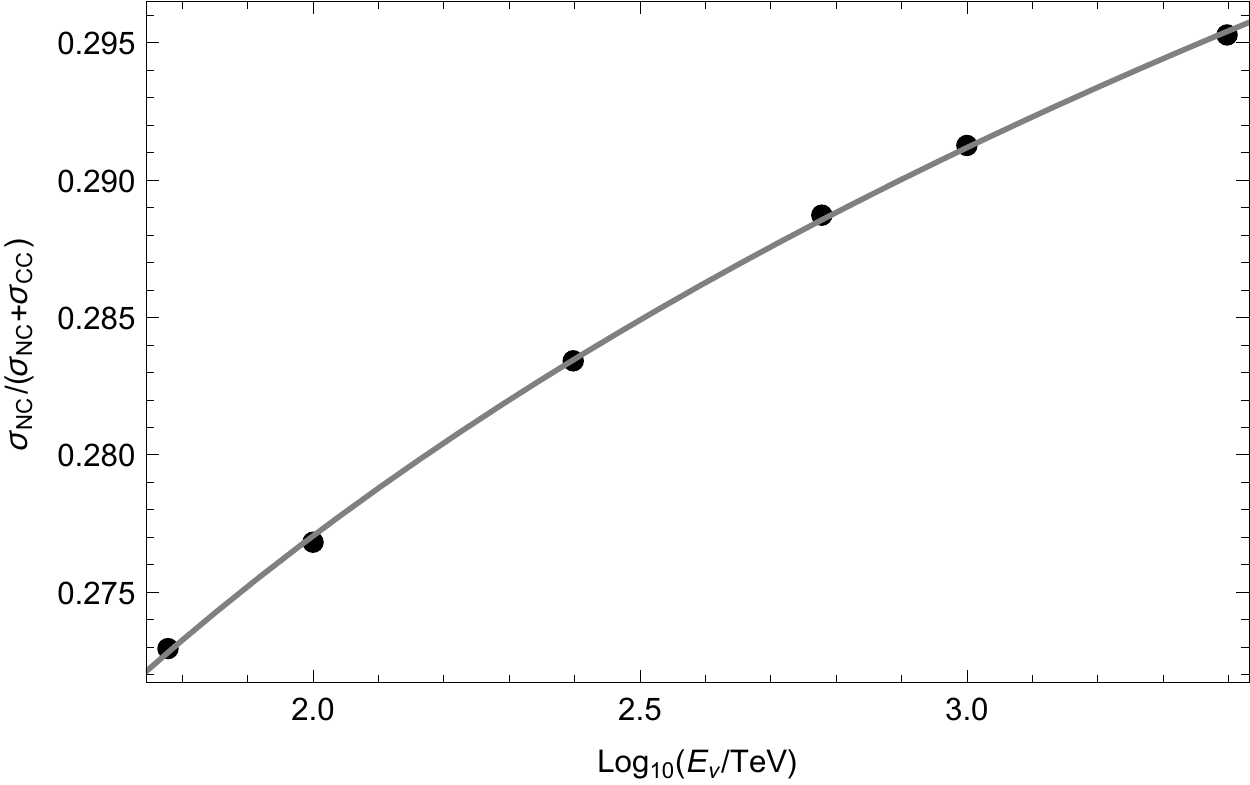}
\caption{Fraction of NC events. Points are taken from Table \ref{tab:sigma_CC} and \ref{tab:sigma_NC} using Eq. \ref{eq:NC_frac}, while the curve is obtained from Eq. \ref{eq:NC_prior}. }
\label{fig:fit_NC}
\end{figure}

Our current knowledge of the proton's parton distributions allows
us to calculate the neutrino-nucleon cross sections with confidence up to neutrino energies of about 10 PeV \cite{Gandhi:1995tf}. At neutrino energies $E_{\nu}$ above some 10 GeV, as relevant for this analysis, neutrino-nucleon reactions are dominated by deep-inelastic scattering. The processes that go into our evaluation are the CC channel, where the $\nu$ scatters off a quark in the nucleon N via exchange of a virtual W-boson,
$$
\nu_{\ell} N \rightarrow X + \ell
$$
and the  NC channel, via exchange of a virtual Z-boson,
$$
\nu_{\ell} N \rightarrow X + \nu_{\ell} ,
$$
where $\ell = \{ e, \mu, \tau \}$ , and
$X$ represents hadrons. In Fig. \ref{fig:CC_NC_plot} both interactions are schematically represented.
The neutrino-nucleon CC and NC cross-sections have been measured by several experiments. A complete review can be found in Ref. \cite{Gandhi:1998ri}, from which we report in Table \ref{tab:sigma_CC} and \ref{tab:sigma_NC} the values of cross-sections respectively for CC and NC interaction for given energy values in the range 60 TeV-3 PeV.
\def\arraystretch{1.5}
\begingroup
 \squeezetable
\begin{table}[!h]
\caption{Charged-current cross sections for neutrino, anti-neutrino and their sum for neutrino-nucleon interactions.}
 \begin{ruledtabular}
\begin{tabular}{lccc}
$E_{\nu}$ [TeV] & $\sigma_{CC}^{\nu}$ [$10^{-33} \, cm^2$] &  $\sigma_{CC}^{\bar{\nu}}$ [$ 10^{-33} \, cm^2$] & $\sigma_{CC}$ [$10^{-33} \, cm^2$]  \\ \hline
60   & 0.1514 & 0.1199 & 0.2713 \\
100 & 0.2022 & 0.1683 &  0.3705 \\
250 & 0.3255 & 0.2909 & 0.6164 \\
600 & 0.4985&  0.4667 & 0.9652 \\
 $10^3$  & 0.6342 & 0.6051 & 1.2393 \\
 $2.5 \cdot 10^3$ & 0.9601 & 0.9365 & 1.8966 \\
\end{tabular}
 \end{ruledtabular}
\label{tab:sigma_CC}
\end{table}
\endgroup
\def\arraystretch{1.5}
\begingroup
 \squeezetable
\begin{table}[!h]
\caption{Neutral-current cross sections for neutrino, anti-neutrino and their sum for neutrino-nucleon interactions.}
 \begin{ruledtabular}
\begin{tabular}{lccc}
$E_{\nu}$ [TeV] & $\sigma_{NC}^{\nu}$ [$10^{-33} \, cm^2$] &  $\sigma_{NC}^{\bar{\nu}}$ [$ 10^{-33} \, cm^2$] & $\sigma_{NC}$ [$ 10^{-33} \, cm^2$]  \\ \hline
60   & 0.05615 & 0.04570 & 0.10185 \\
100 & 0.07667 & 0.06515 & 0.14182  \\
250 & 0.1280 & 0.1158 & 0.2438 \\
600 & 0.2017 &  0.1901 & 0.3918 \\
 $10^3$  & 0.2600 & 0.2493 & 0.5093 \\
 $2.5 \cdot 10^3$ & 0.4018 & 0.3929 & 0.7947 \\
\end{tabular}
\end{ruledtabular}
\label{tab:sigma_NC}
\end{table}
\endgroup
For the purpose of this analysis, we need to know the probability that a neutrino interacts with a nucleon via CC or NC channel. In order to estimate this probability we use the values given in Table \ref{tab:sigma_CC} and \ref{tab:sigma_NC} from which we get the fraction of NC events
\begin{equation}
\frac{\sigma_{NC}}{\sigma_{CC}+ \sigma_{NC} }.
\label{eq:NC_frac}
\end{equation}
These values are then fitted, as shown in Fig. \ref{fig:fit_NC}, in order to obtain the following parametrization in terms of $\epsilon = \text{Log}_{10}( E_{\nu} / \text{TeV})$
\begin{equation}
A_1 + A_2 \cdot \log (\epsilon - A_3 ),
\label{eq:NC_prior}
\end{equation}
with $A_1 = 0.2595$, $A_2 = 0.0313$ and $A_3 = 0.2484$. From Fig. \ref{fig:fit_NC} one can see that the probability that a neutrino interacts with a nucleon via NC is $\sim 30 \%$ and in the range of energies we are interested in depends slightly on the neutrino energy $E_{\nu}$. Eq. \ref{eq:NC_prior} will be then used as the prior probability for NC interactions.

An important parameter that plays a crucial role in this analysis is the inelasticity parameter $y$: as schematically shown in Fig. \ref{fig:CC_NC_plot}, in both CC and NC interactions a fraction ($1-y$) of the neutrino energy $E_{\nu}$ goes to the final-state lepton; the remaining fraction $y$ goes to the final-state hadrons. 

The differential cross section for CC interactions in terms of $y$ and of the Bjorken scaling variables $x$ (the fraction of the nucleon momentum
carried by the struck quark) is given by (in natural units $ \hslash = c= 1$)
\begin{equation}
\frac{d \sigma_{CC}}{dy dx} =\frac{2 G_F^2 M_N E_{\nu}}{\pi} \left( \frac{M_{W}^2}{Q^2+M_W^2} \right)^2 \left( q + (1-y)^2\bar{q} \right).
\label{eq:CC_cros}
\end{equation}
Likewise, the NC differential cross section is given by
\begin{equation}
\frac{d \sigma_{NC}}{dy dx} =\frac{2 G_F^2 M_N E_{\nu}}{\pi} \left( \frac{M_{Z}^2}{Q^2+M_Z^2} \right)^2 \left( q^0 + (1-y)^2\bar{q}^0 \right).
\label{eq:NC_cros}
\end{equation}
In these equations $q$, $\bar{q}$, $ q^0$ and $\bar{q}^0$ are quark and antiquark distribution functions \cite{Connolly:2011vc, Gandhi:1998ri}, $M_N$, $M_W$ and $M_Z$ are respectively the nucleon, W and Z mass, $G_F$ is the Fermi coupling constant and $Q^2 \approx 2xyE_{\nu}M_N$ is the negative four-momentum transfer squared.

In order to simulate in our code the $y$-distribution given by Eq. \ref{eq:CC_cros}  and \ref{eq:NC_cros}, we used the algorithm described in Ref. \cite{Connolly:2011vc}.

\begin{figure}[!h]
\includegraphics[width=0.95\linewidth,height=0.17\textheight]{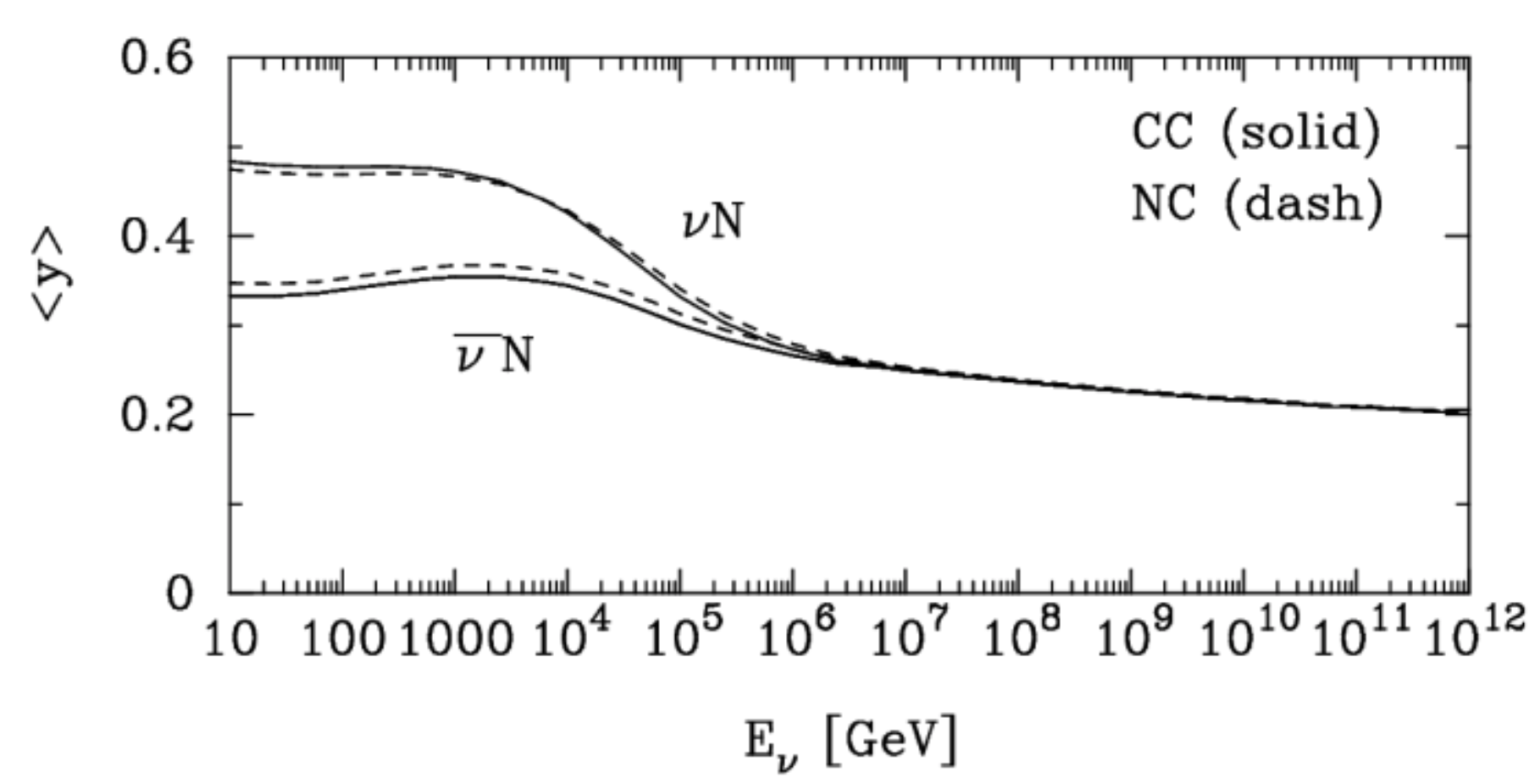}
\caption{Average $y$ as a function of
neutrino energy $E_{\nu}$, for CC (solid lines) and
NC (dashed) reactions. Figure taken
from Ref. \cite{Gandhi:1995tf}. }
\end{figure}

\section{\label{sec:tau_decay} $\tau$-decay channels}

\begin{figure}[!h]
\begin{tikzpicture}
\begin{feynman}
\vertex (a) {\(\tau^{-}\)};
\vertex [right=1 cm of a] (b);
\vertex [above right=1 cm of b] (f1) {\(\nu_{\tau}\)};
\vertex [below right=1 cm of b] (c);
\vertex [above right=1 cm of c] (f2) {\(\overline \nu_{e} , \overline \nu_{\mu}\)};
\vertex [below right=1 cmof c] (f3) {\(e^{-}, \mu^{-} \)};
\diagram* {
(a) -- [fermion] (b) -- [fermion] (f1),
(b) -- [boson, edge label'=\(W^{-}\)] (c),
(c) -- [anti fermion] (f2),
(c) -- [fermion] (f3),
};
\end{feynman}
\end{tikzpicture}
%
%$\qquad$
%
\begin{tikzpicture}
\begin{feynman}
\vertex (a) {\(\tau^{-}\)};
\vertex [right=1 cm of a] (b);
\vertex [above right=1 cm of b] (f1) {\(\nu_{\tau}\)};
\vertex [below right=1 cm of b] (c);
\vertex [ right=1 cm of c] (f2) ;
\vertex [ right=1.5 cm of f2] (f3) ;
\diagram* {
(a) -- [fermion] (b) -- [fermion] (f1),
(b) -- [boson, edge label'=\(W^{-}\)] (c),
(c) -- [anti fermion, half left, edge label=\( \bar{u}\)] (f2),
(c) -- [ fermion, half right, edge label=\( d \)] (f2),
(f2) -- [ fermion, very thick, edge label= hadrons] (f3),
};
\end{feynman}
\end{tikzpicture}
 \caption{
        Diagrams for leptonic (left) and hadronic (right) decay of the $\tau$ lepton. Time runs from left to right.
    }
    \label{fig:tau_decay_plot}
\end{figure}
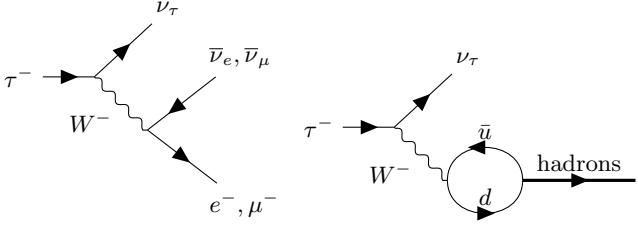

When a $\nu_{\tau}$ and a nucleon interacts via CC interaction a $\tau$ of energy $E_{\tau} = (1-y) E_{\nu}$ is produced. The $\tau$ is the heaviest of the leptons with a mass $m_{\tau}$ of 1.78 GeV and therefore it has a very short lifetime of about $3 \cdot 10^{-13} \, s$.  It can decay in the lepton channel or in the hadronic channel, as shown schematically in Fig. \ref{fig:tau_decay_plot}. The leptonic decays have a total branching fraction of $\sim 35 \%$ and the hadronic decays have a total branching fraction of $\sim 65 \%$, which is consistent with the expected branching fraction when the color charges of the quarks are included. 
The branching fraction into each decay channel is approximately \cite{Dutta:2000jv}
\begin{align}
\begin{split}
0.18 \quad  & \text{for} \; \tau \rightarrow \nu_{\tau} e \nu_e, \\
0.18 \quad & \text{for} \; \tau \rightarrow \nu_{\tau} \mu \nu_{\mu},  \\
0.12 \quad  & \text{for} \; \tau  \rightarrow \nu_{\tau} \pi ,\\
0.26  \quad & \text{for} \; \tau  \rightarrow \nu_{\tau} \rho , \\
0.13 \quad  & \text{for} \; \tau  \rightarrow \nu_{\tau} a_1, \\
0.13 \quad & \text{for} \; \tau  \rightarrow \nu_{\tau} X  \quad ( X \neq\pi , \rho, a_1).   \\
\end{split}
\label{eq:br_fraction}
\end{align}
Due to its very short lifetime the track produced inside the detector by the $\tau$  has generally a length of
$
50 \, m \cdot \left( E_{\tau} / \text{PeV} \right)
$
\cite{Xu:2017yxo}. At energies below PeV, the double cascade signature is difficult to distinguish from a single cascade, due to the sparse spacing of digital optical modules. Thus a track produced by a $\tau$ below a few PeV is unresolvable by IceCube. At higher energy ($ \gtrsim 1$ PeV) a signature of $\nu_{\tau}$ CC interactions would be two cascades joined by a short track, referred as a "double bang", which has not yet been observed.  Considering the energies of our interest, in this analysis we assume that $\nu_{\tau}$ CC interactions followed by $\tau \rightarrow \nu_{\tau} \mu \nu_{\mu}$ are undistinguishable  from a track event produced in $\nu_{\mu}$ CC interactions, while all the other $\tau$-decay channels produce a shower event.

The $\nu_{\tau}$ spectra for $\tau$-leptonic decay has the following form in term of $z= E_{\nu_{\tau}} / E_{\tau}$ \cite{LIPARI1993195}
\begin{equation}
\frac{d\sigma}{dz} \propto  \left(\frac{5}{3} -3 z^2+ \frac{4 z^3}{3} \right) - P_{\tau} \left( \frac{1}{3}-3 z^2+ \frac{8 z^3}{3} \right) ,
\label{eq:z_lepton}
\end{equation}
while the $\nu_{\ell}$ ($\ell = \{e , \mu \}$) spectra in term of $z'= E_{\nu_{\ell}} / E_{\tau}$ reads
\begin{equation}
\frac{d\sigma}{dz'} \propto  \left(2 -6 {z'}^2+4 z'^3 \right) - P_{\tau} \left(-2+ 12 z' -18 z'^2+8 z'^3 \right) ,
\label{eq:zp_lepton}
\end{equation}
where $P_{\tau}$ is the polarization of the $\tau$. 

In the case of hadronic decays $ \tau  \rightarrow \nu_{\tau} X$ the distribution depends on the kind of hadrons produced. An approximation of the distribution for each hadronic channel $i$, in terms  of $z= E_{\nu_{\tau}} / E_{\tau}$ and $r_i = m_i^2/m_{\tau}^2$, can be found in Ref. \cite{Dutta:2000jv}:
\begin{widetext}
\begin{equation}
\frac{d\sigma}{dz} \propto
\begin{cases}
\frac{1}{1-r_{\pi}} \theta(1- r_{\pi} - z) + P_{\tau} \frac{2z -1 + r_{\pi}}{(1-r_{\pi})^2} \theta(1- r_{\pi} - z), &  \tau  \rightarrow \nu_{\tau} \pi ,\\
\frac{1}{1-r_{\rho}} \theta(1- r_{\rho} - z) + P_{\tau} \left( \frac{2z -1 + r_{\rho}}{1-r_{\rho}} \right) \left( \frac{1 - 2 r_{\rho}}{1+ 2 r_{\rho}} \right)  \theta(1- r_{\rho} - z), &  \tau  \rightarrow \nu_{\tau} \rho, \\
\frac{1}{1-r_{a_1}} \theta(1- r_{a_1} - z) + P_{\tau} \left( \frac{2z -1 + r_{a_1}}{1-r_{a_1}} \right) \left( \frac{1 - 2 r_{a_1}}{1+ 2 r_{a_1}} \right)  \theta(1- r_{a_1} - z), &  \tau  \rightarrow \nu_{\tau} a1, \\
\frac{1}{0.3} \theta(0.3-z), & \tau  \rightarrow \nu_{\tau} X  \quad ( X \neq\pi , \rho, a1).
\end{cases}
\label{eq:z_hadron}
\end{equation}
\end{widetext}
For energies of our interest we have $m_{\tau}/E_{\tau} \ll 1$, thus it is safe to assume \cite{Hagiwara:2003di,Bourrely:2004iy} the $\tau$ being almost fully polarized, i.e., $P_{\tau} = 1$. 

The distributions in Eq. \ref{eq:z_lepton}, \ref{eq:zp_lepton} and \ref{eq:z_hadron} , along with their respective branching fraction in Eq. \ref{eq:br_fraction}, will be then used as prior distributions in those CC interactions involving a $\nu_{\tau}$ and its subsequent decay.

\section{\label{sec:deposited_energy} Deposited energy}

All charged particles produced in the nucleon-neutrino interaction propagate through ice emitting Cherenkov radiation. This Cherenkov radiation is ultimately measured by the IceCube detectors producing a deposited energy $E_{dep.}$, which is proportional to the total energy $E_{\nu}$ of the neutrino.
Each channel has different efficiencies when it comes to producing a measured energy deposition in the IceCube detector.
First of all one has to distinguish electromagnetic cascades from hadronic cascades, which are both recognize in the detector as showers. For electromagnetic cascades one can safely assume  the deposited energy being equal to the energy of the electron produced in the nucleon-neutrino interaction.  On the other hand, the deposited energy in hadronic cascade is less reliable due to the presence of more neutral particles like neutrons, to large losses due to the binding energies in hadronic processes and to a higher Cherenkov threshold for hadrons \cite{Kowalski2004Search}.

Following Ref. \cite{Palomares-Ruiz:2015mka}, being $E_X$ the energy of the cascade-initiating particle, we define the deposited energy in hadronic cascade as
\begin{equation}
E_h (E_X) =  \left( 1 - f \cdot \left( \frac{E_X}{E_0} \right)^{-m} \right) \cdot E_X,
\end{equation}
where $f = 0.533$, $E_0 = 0.399 \, GeV$ and $ m = 0.130$, resulting from a fit to simulations of hadronic cascades \cite{Kowalski2004Search}.
 
For track events, being the lifetime of a muon much larger than the time it takes to cross the detector, a fraction of the initial muon energy $E_{\mu}$ is lost. The average deposited energy along a track $E_t$ by a muon can be obtained using the parametrization given in  Ref. \cite{Palomares-Ruiz:2015mka}
\begin{equation}
E_{t} (E_{\mu} ) = F_{\mu} \cdot ( E_{\mu} + a/b),
\end{equation}
where $a=0.206 \, GeV/m$,  $b=3.21 \cdot 10^{-4} \, m^{-1}$ and $F_{\mu} = 0.119$.
If the track is produced in a tau decay of energy $E_{\tau}$, one has to take into account also that a significant fraction of tau leptons would escape the detector volume before decaying, so that $F_{\mu} $ has to be multiplied by a factor given by 
\begin{equation}
 \frac{1 + p_1 \cdot (E_{\tau}/10 \, PeV)}{1 + q_1 \cdot (E_{\tau}/10 \, PeV) + q_2 \cdot (E_{\tau}/10 \, PeV)^2},
\end{equation}
where $p_1 = 0.984 $, $q_1 = 1.01 $ and $ q_2 = 1.03 $ \cite{Palomares-Ruiz:2015mka}. 

Finally, for all the nucleon-neutrino interaction we have considered, the total deposited energy $E_{dep.}$ is given by
\begin{widetext}
\begin{equation}
 E_{dep.} = 
		\begin{cases}   		
   		E_h ( E_X) , &   \quad NC,  \\
  		E_h ( E_X) +E_{\ell}  , &   \nu_e \, CC,  \\
  		E_h ( E_X) + E_t(E_{\ell})  , &  \nu_{\mu} \, CC,  \\
  		E_h ( E_X) +E_{\ell} \cdot (1 - z-z') , &  \nu_{\tau}  \,  CC \quad \tau \rightarrow \nu_{\tau} e \nu_e, \\
  		E_h ( E_X) +E_t \left( E_{\ell} \cdot (1 - z-z') \right) , &  \nu_{\tau} \,  CC \quad \tau \rightarrow \nu_{\tau} \mu \nu_{\mu},  \\
  		E_h ( E_X) + E_h \left( E_{\ell} \cdot (1-z) \right), &   \nu_{\tau} \,  CC \quad \tau \rightarrow\nu_{\tau} X, \\
   		\end{cases}
 \label{eq:en_dep} 
\end{equation}
\end{widetext}
where $E_X = y  E_{\nu}$,  $E_{\ell} =  (1-y) E_{\nu}$ with $\ell = \{e, \mu, \tau \}$, while $y$ has been discussed in Sec. \ref{sec:scattering}. The parameters $z  $ and $z' $ have been discussed in Sec. \ref{sec:tau_decay}, which are respectively  $E_{\nu_{\tau}}/ E_{\tau}$ and $E_{\nu_{e,\mu}}/E_{\tau}$.

For both topologies the energy deposited within the detector can be reconstructed within $\sim 15 \%$ above 10 TeV \cite{Aartsen:2013vja}. Thus one has to make distinction between the true deposited energy $E_{dep.}$ and the observed-deposited energy $E_{dep.}^{obs.}$. For this analysis we simply assume that  $E_{dep.}^{obs.}$ follows a normal distribution with mean value given by $E_{dep.}$ and standard deviation $\sigma_{E_{dep.}}$:
\begin{align}
\begin{split}
\mathcal{N} & (E_{dep.}^{obs.} \, | \, E_{dep.},\sigma_{E_{dep.}}) = \\
= & \frac{1}{{\sigma_{E_{dep.}} \sqrt {2\pi } }}e^{{{ - \left( {E_{dep.}^{obs.} - E_{dep.} } \right)^2 } \mathord{\left/ {\vphantom {{ - \left( {x - \mu } \right)^2 } {2\sigma_{E_{dep.}} ^2 }}} \right. \kern-\nulldelimiterspace} {2\sigma_{E_{dep.}}^2 }}}  \text{.}
\end{split}
\end{align}
For each neutrino event the value of $\sigma_{E_{dep.}}$ is taken from the uncertainty in the deposited energy provided by IceCube \cite{Aartsen:2015zva,Aartsen:2017mau}.

\section{\label{sec:analysis}Analysis}

In Table \ref{tab:parameters} we summarize all parameters and the sequence of events that, given a neutrino with energy $E_{\nu}$, cause an observed-deposited energy $E_{dep.}^{obs.}$ in the detector. In a certain sense, we need to go backwards through the whole chain of events in order to infer the neutrino energy $E_{\nu}$ from the observed-deposited energy $E_{dep.}^{obs.}$ and its topology (which sometimes in the rest of this paper we abbreviate as ''top.''). In this section we briefly describe how this goal can be achieved using Bayesian inference.

\def\arraystretch{1.7}
\begingroup
 \squeezetable
\begin{table*}
    \caption{ Table of all parameters used in this analysis along with their associated prior probability distribution.  The right two columns show the sections and the references where these parameters are discussed in detail. The values of $a$ and $b$ in the $z$-parameter row can be obtained from Eq. \ref{eq:z_hadron} with $P_{\tau} = 1$.}
    \begin{ruledtabular}
    \begin{tabular}{llcc}
        Parameters & Prior probability distribution & Sec. & Ref. \\
        \hline \multicolumn{4}{c}{Flux parameters} \\ \hline
       % & &  &  \\
       $r \quad$ (anti-neutrino/neutrino ratio) & $\delta( r -1)$ & \ref{sec:Flux}  & \cite{Nunokawa:2016pop}  \\
       % & & & \\
		$\ell \quad$  (neutrino flavor) & $\begin{aligned}
1/3, \quad & \ell = e \\
1/3, \quad & \ell = \mu \\
1/3, \quad & \ell = \tau 
\end{aligned}$   & \ref{sec:Flux} & \cite{Aartsen:2015ivb}  \\
		% & & & \\
		$ \gamma \quad$ (spectral index) & $\mathcal{N}(\gamma \, | \, 2.13, 0.13) $ & \ref{sec:Flux}    & \cite{Aartsen:2016xlq} \\
		% & &  &   \\
        $E_{\nu} \quad$ (neutrino energy) & $(\gamma-1) \, E_{\nu}^{-\gamma}/ \left( (60 \, \text{TeV})^{-\gamma+1} - (3 \, \text{PeV})^{-\gamma+1} \right)  $  & \ref{sec:Flux}  &  \cite{Aartsen:2016xlq} \\
        % & &  &  \\
        \hline \multicolumn{4}{c}{Deep-inelastic scattering parameters} \\ \hline
        $k \quad$ (nucleon-neutrino interaction)  & $\begin{aligned}  
   		A_1 + A_2 \cdot \ln( \epsilon- A_3), \quad &  k = \text{NC} \\
  		1- A_1 - A_2 \cdot \ln( \epsilon- A_3), \quad  & k = \text{CC} \\
   		\end{aligned}$ & \ref{sec:scattering}   & \cite{Connolly:2011vc, Gandhi:1995tf} \\
   		% & &  &  \\
        $y \quad$ (inelasticity parameter) &  $d \sigma_k ( E_{\nu})/d y  \quad$ (see Eq. \ref{eq:CC_cros} and \ref{eq:NC_cros}) &  \ref{sec:scattering} & \cite{Connolly:2011vc,Gandhi:1995tf} \\
       % & &  &  \\
        \hline \multicolumn{4}{c}{$\tau$-decay parameters} \\ \hline
        %& &  &  \\
        $j \quad$ ($\tau$-decay channel) & 
       $\begin{aligned}
0.18, \quad  &  j= \tau \rightarrow \nu_{\tau} e \nu_e \\
0.18, \quad & j=  \tau \rightarrow \nu_{\tau} \mu \nu_{\mu}  \\
0.12, \quad  & j=  \tau  \rightarrow \nu_{\tau} \pi \\
0.26,  \quad & j=  \tau  \rightarrow \nu_{\tau} \rho  \\
0.13, \quad  & j=  \tau  \rightarrow \nu_{\tau} a_1 \\
0.13, \quad & j=  \tau  \rightarrow \nu_{\tau} X  \quad ( X \neq\pi , \rho, a_1)
\end{aligned}$
         & \ref{sec:tau_decay}  & \cite{Dutta:2000jv} \\
         %& &  &  \\
        $ z \quad$ (energy fraction $E_{\nu_{\tau}}/E_{\tau}$) & 
        $\begin{aligned}
 4/3 \left( 1 - z^3 \right), \quad   & \text{if  }  j= \tau \rightarrow \nu_{\tau} e \nu_e \; \text{or} \;\nu_{\tau} \mu \nu_{\mu} \\
 \left(a_{\pi}  +  b_{\pi}  \cdot z \right)  \theta(1- r_{\pi} - z) , \quad  & \text{if  } j=  \tau  \rightarrow \nu_{\tau} \pi \\
\left(a_{\rho}  +  b_{\rho}  \cdot z \right)  \theta(1- r_{\rho} - z), \quad  &  \text{if  }  j=  \tau  \rightarrow \nu_{\tau} \rho  \\
\left(a_{a_1}  +  b_{a_1}  \cdot z \right)  \theta(1- r_{a_1} - z),  \quad  & \text{if  }  j=  \tau  \rightarrow \nu_{\tau} a_1 \\
1 / 0.3 \, \theta(0.3-z),  \quad & \text{if  }  j=  \tau  \rightarrow \nu_{\tau} X  \quad ( X \neq\pi , \rho, a_1)
\end{aligned}$
         & \ref{sec:tau_decay}  & \cite{Hagiwara:2003di, LIPARI1993195, Dutta:2000jv} \\
         %
         % & &  &  \\
        $ z' \quad$ (energy fraction $E_{\ell}/E_{\tau}$) & $ 4 - 12 z' + 12 z'^2 - 4 z'^3, \quad \text{if  }  j= \tau \rightarrow \nu_{\tau} e \nu_e \; \text{or} \; \nu_{\tau} \mu \nu_{\mu}$ & \ref{sec:tau_decay}  & \cite{LIPARI1993195, Dutta:2000jv} \\
        \hline \multicolumn{4}{c}{Deposited Energy} \\ \hline
         $E_{dep.}^{obs.} \;$ (observed deposited energy) & 
         $\; \mathcal{N}(E_{dep.}^{obs.} \, | \, E_{dep.},\sigma_{E_{dep.}}) \quad $
        with $  E_{dep.}$ defined in  Eq. \ref{eq:en_dep}
        & \ref{sec:deposited_energy}  & \cite{Palomares-Ruiz:2015mka, Aartsen:2013vja} \\
        %& &  &  \\
    \end{tabular}
    \end{ruledtabular}
    \label{tab:parameters}
\end{table*}
\endgroup

As usually done in literature, let $D$ denote the observed data, in our case the deposited energy $E_{dep.}$ and the event topology (track or shower), and $\theta$ denote the model parameters, which are summarized in the first column of Table \ref{tab:parameters}. Formal inference then requires setting up a joint probability distribution $f(D, \theta)$ (here and in the rest of this paper we will refer simply as $f$ to all distributions). This joint distribution comprises two parts: a prior distribution $f(\theta)$ (see the second column of Table \ref{tab:parameters})  and a likelihood $f(D | \theta)$. Defining $f(\theta)$ and $f(D | \theta)$ gives the full probability distribution
\begin{equation}
f(D, \theta) = f(D | \theta) \cdot  f(\theta).
\end{equation}
Having observed $D$, one can then obtain the distribution of $\theta$ conditional on $D$ by applying the Bayes theorem
\begin{equation}
f(\theta | D) = \frac{f(D | \theta) \cdot  f(\theta)}{\int f(D | \theta) \cdot  f(\theta) \, d \theta}.
\label{eq:bayes}
\end{equation}
This is called the posterior distribution of $\theta$ and is the object of our Bayesian-inference analysis. From the posterior distribution of $\theta$ one can then obtain the expected value of a given parameter by integrating over the remaining parameters or study the dependence between parameters $x$ and $y$ by applying the product rule $f(x| y , D) = f(x , y| D) / f(y | D)$.

From Eq. \ref{eq:bayes}, one recovers the maximum likelihood approach as a special case that holds under particular conditions, such as many data points and vague priors, which clearly are not satisfied in this analysis.

In theory, Bayesian methods are straightforward: the posterior distribution contains everything you need to carry out inference.
In practice, the posterior distribution can be difficult to estimate precisely.
A useful tool to derive the posterior distribution of Eq. \ref{eq:bayes} is the Markov Chain Monte Carlo (MCMC) technique. In a MCMC instead of having each point being
generated one independently from another (like in a Monte Carlo), the sequence of generated points takes a kind of random walk in parameter space. Moreover, the probability of jumping from one
point to an other depends only on the last point and not on the entire previous history (this is the peculiar property of a Markov chain). In particular, for this work we performed the MCMC using the Gibbs sampling algorithm \cite{doi:10.1080/01621459.2000.10474335}, in order to explore the entire parameter space of the posterior distribution. This allows us to derive the unknown and potentially complex distribution $f(\theta | D)$ and estimate all neutrino properties we are interested in. The results of this inference analysis are presented and discussed in Sec. \ref{sec:results}.

\section{\label{sec:results}Results and Conclusion}

In Table \ref{tab:results} we show for each of the 37 shower events above 60 TeV, denoted by its ID number and observed-deposited energy $E_{dep.}^{obs.}$, the mean values (mean) and  the standard deviations (s.d.) of the posterior distribution of neutrino energy $E_{\nu}$. The mean and s.d. values are given assuming different flavors $\ell$ ($e$, $\mu$ or $\tau$) and type of interaction $k$ (CC or NC), where for the meaning of parameters $\ell$ and $k$ we remind the reader to see Table \ref{tab:parameters}. In the last columns, one can also find for each neutrino the probability $f( \ell | E_{dep.}^{obs.}, \text{top.} )$ of being of electronic, muonic and tauonic flavor and the probability $f(k | E_{dep.}^{obs.}, \text{top.} )$ of having scattered with nucleon via CC or NC interaction. 
We show the same results for the 13 track events above 60 TeV in Table \ref{tab:reasults_track}. But in this case the probabilities for neutrinos of being electronic or having scattered with nucleon via CC or NC interaction are absent: as we learned in the previous sections, tracks can only be produced in CC interactions by muonic or tauonic neutrinos.

\begin{figure*}
    \includegraphics[width=0.49\linewidth,height=0.35\textheight]{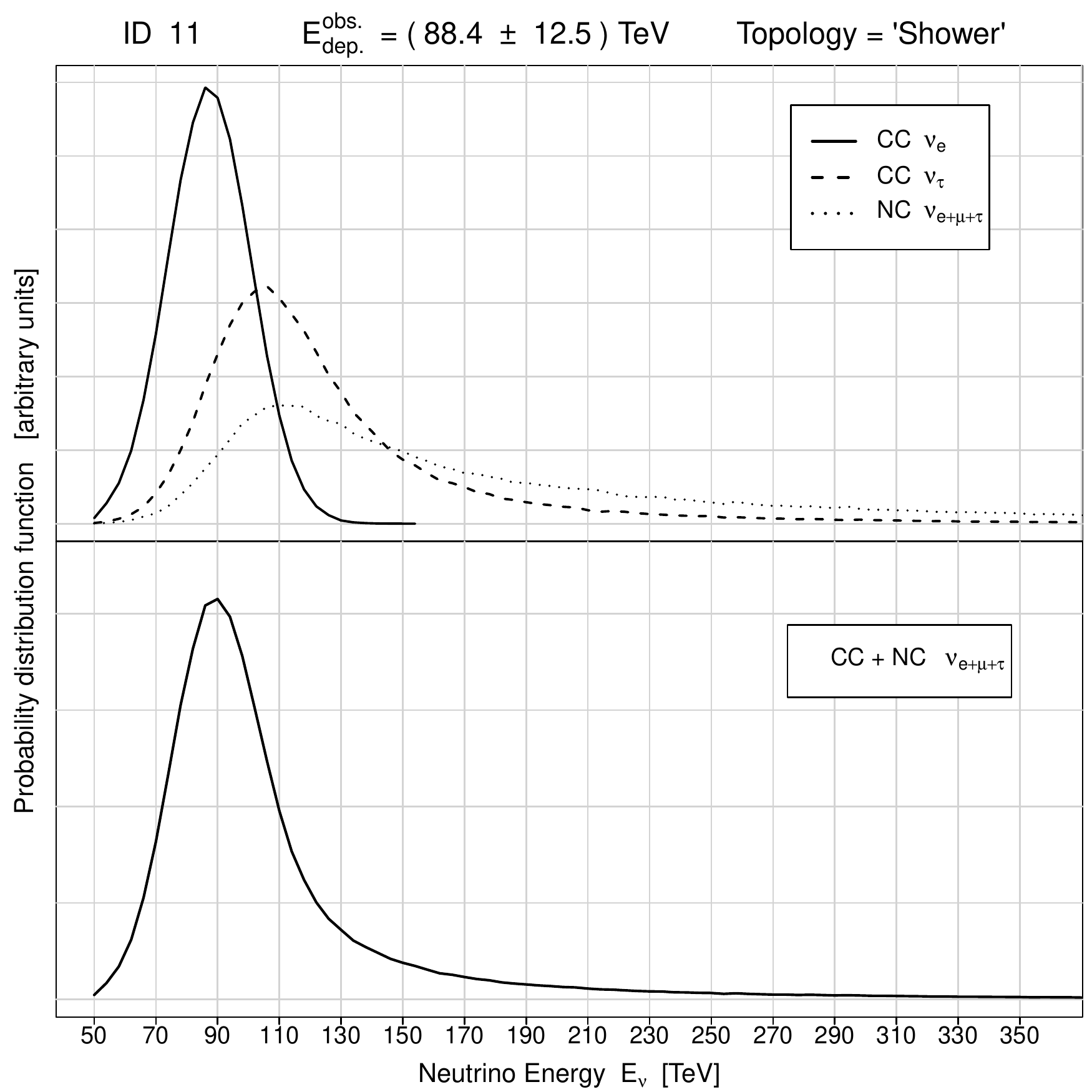}
    \includegraphics[width=0.49\linewidth,height=0.35\textheight]{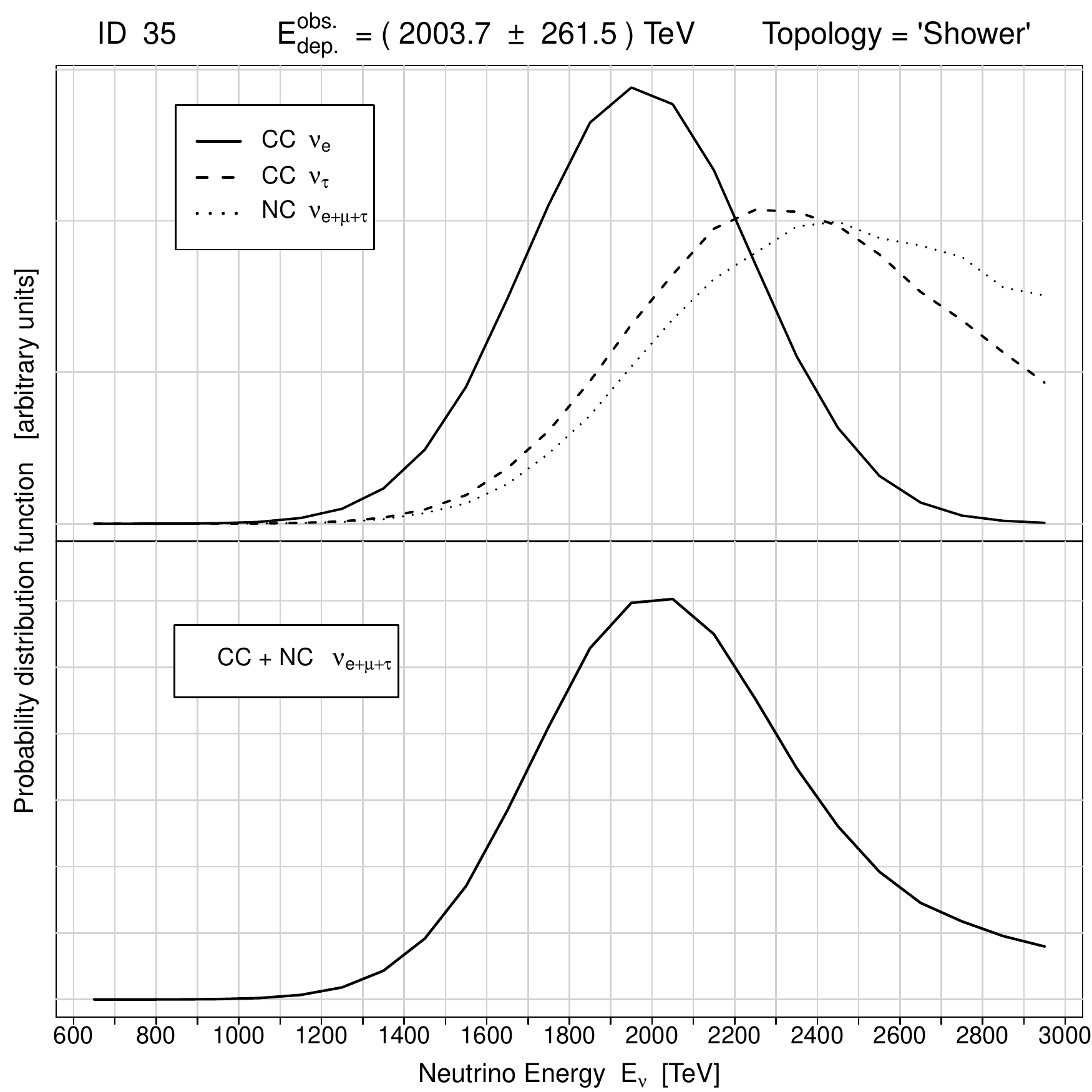} 
\caption{Posterior probability distributions $f( E_{\nu} | \ell, k, E_{dep.}^{obs.}, \text{shower})$ of the neutrino energy for two shower events. In the top panel the distributions assuming $\nu_e$ CC interaction (solid line), $\nu_{\tau}$ CC interaction (dashed line) and NC interaction (dotted line) are shown. In the bottom panel the distribution $f( E_{\nu} |  E_{dep.}^{obs.}, \text{shower})$, obtained marginalizing over $\ell$ and $k$,  is shown.}
\label{fig:plots_PDF_shower}
\end{figure*}

For shower events the neutrino energy $E_{\nu}$ is, as expected, approximately equal to the observed-deposited energy $E_{dep.}^{obs.}$ only in $\nu_e$ CC interactions, where the uncertainty (given by the s.d.) for $E_{\nu}$ is also approximately equal to the uncertainty $\sigma_{E_{dep.}}$ in the observed-deposited energy. For $\nu_{\tau}$ CC interactions and all-flavors NC interactions instead the situation is different: due mainly to neutrinos energy loss in neutrino-nucleon deep-inelastic scattering and to the $\tau$-decay products escaping the detector, the neutrino energy results being higher than the observed-deposited energy with a more dispersed distribution. This behaviour is manifest in Fig. \ref{fig:plots_PDF_shower}, where the posterior distribution 
\begin{equation}
f( E_{\nu} | \ell, k, E_{dep.}^{obs.}, \text{top.})
\end{equation}
 are shown for two shower events: with observed-deposited energy $E_{dep.}^{obs.} =(88.4 \pm 12.5) $ TeV and  $E_{dep.}^{obs.} =(2003.7 \pm 261.5) $ TeV (the most energetic event).  In the bottom part of these plots is also shown the neutrino-energetic distribution $f(E_{\nu}  | E_{dep.}^{obs.}, \text{top.})$ making no assumption on $\ell$ and $k$, i.e. marginalizing over these parameters
 \begin{widetext}
\begin{equation}
f(E_{\nu}  | E_{dep.}^{obs.}, \text{top.} ) = \sum_{\ell, k} f(E_{\nu} | \ell, k  , E_{dep.}^{obs.}, \text{top.} ) \cdot f(\ell  | E_{dep.}^{obs.}, \text{top.} ) \cdot f(k  | E_{dep.}^{obs.}, \text{top.} ).
\end{equation}
\end{widetext}

\begin{figure*}
    \includegraphics[width=0.49\linewidth,height=0.3\textheight]{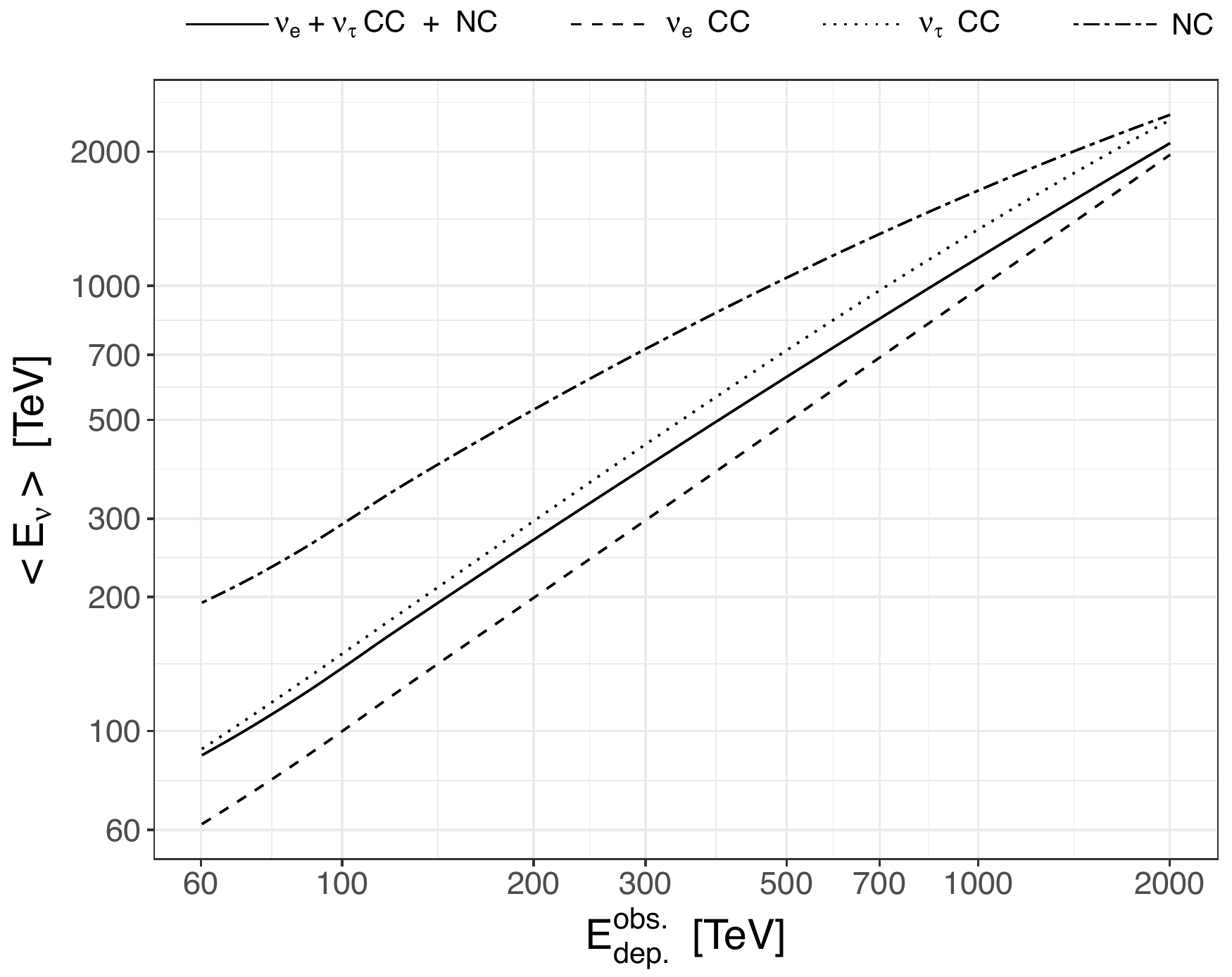}
    \includegraphics[width=0.49\linewidth,height=0.3\textheight]{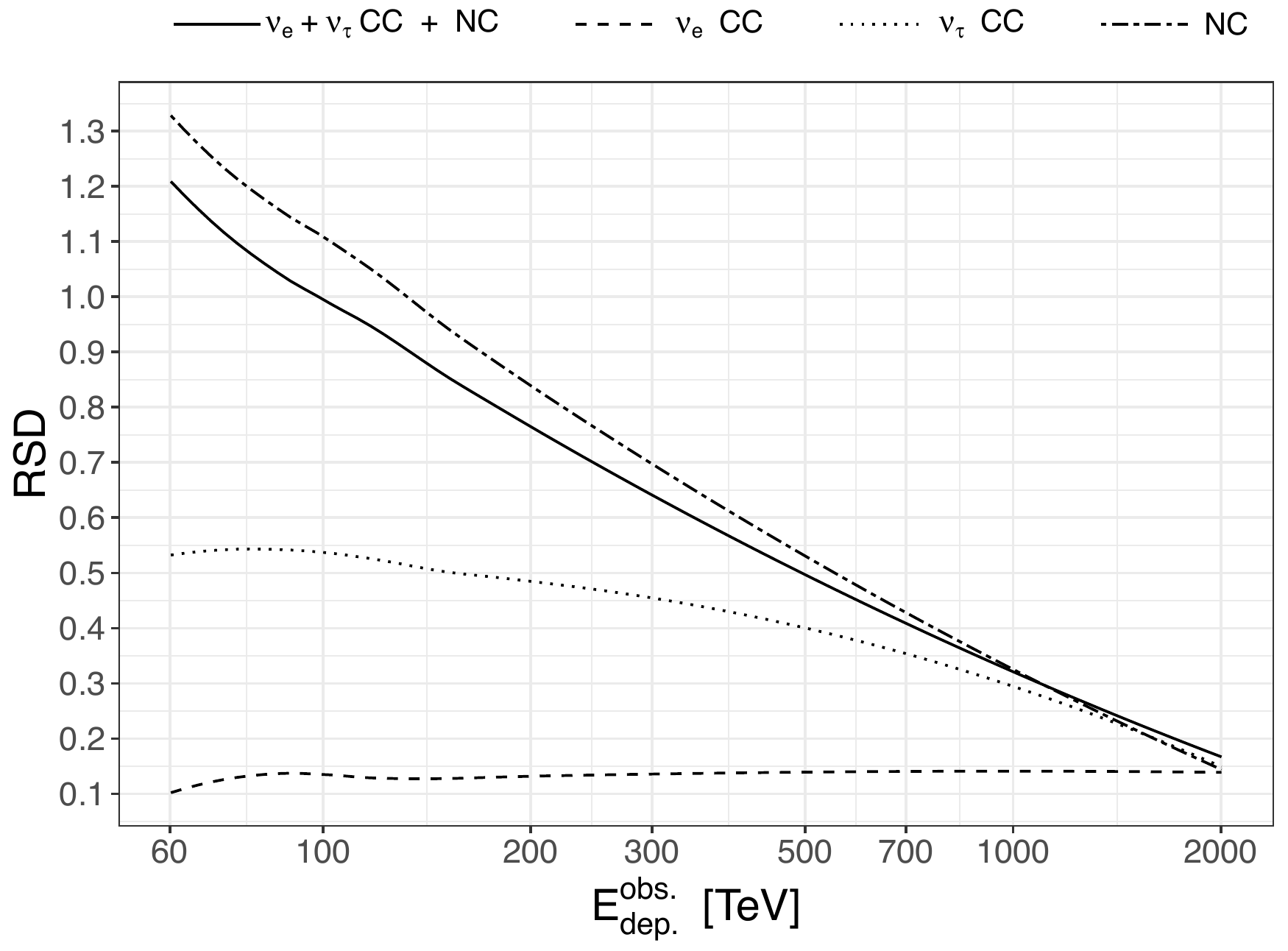}
\caption{The mean value of the neutrino energy $E_{\nu}$ (left) and its RSD value (right) are shown as a function of the observed-deposited energy in shower events making no assumption (solid line), assuming $\nu_e$ CC interaction (dashed line), $\nu_{\tau}$ CC interaction (dotted line) and NC interaction (dot-dashed line).}
\label{fig:plots_E_shower}
\end{figure*}

As one can see from Fig. \ref{fig:plots_PDF_shower}, for showers, having to guess about the neutrino energy, the observed-deposited energy in the detector is the best choice, being this value approximately equal to the mode of the distribution $f(E_{\nu}  | E_{dep.}^{obs.}, \text{top.})$. Instead the mean value feels the effect of the pronounced tail at higher energy produced by NC and $\nu_{\tau}$ CC interactions. Thus the mean value of $E_{\nu}$ results being higher than the observed-deposited energy: in Fig. \ref{fig:plots_E_shower} we show, for different kinds of interaction, the neutrino-energy mean value as a function of the observed-deposited energy and the \emph{relative standard deviation} (RSD), which is a measure of dispersion of a probability distribution (defined as the ratio of the standard deviation to the mean value), as a function of the observed-deposited energy.

\begin{figure*}
    \includegraphics[width=0.48\linewidth,height=0.35\textheight]{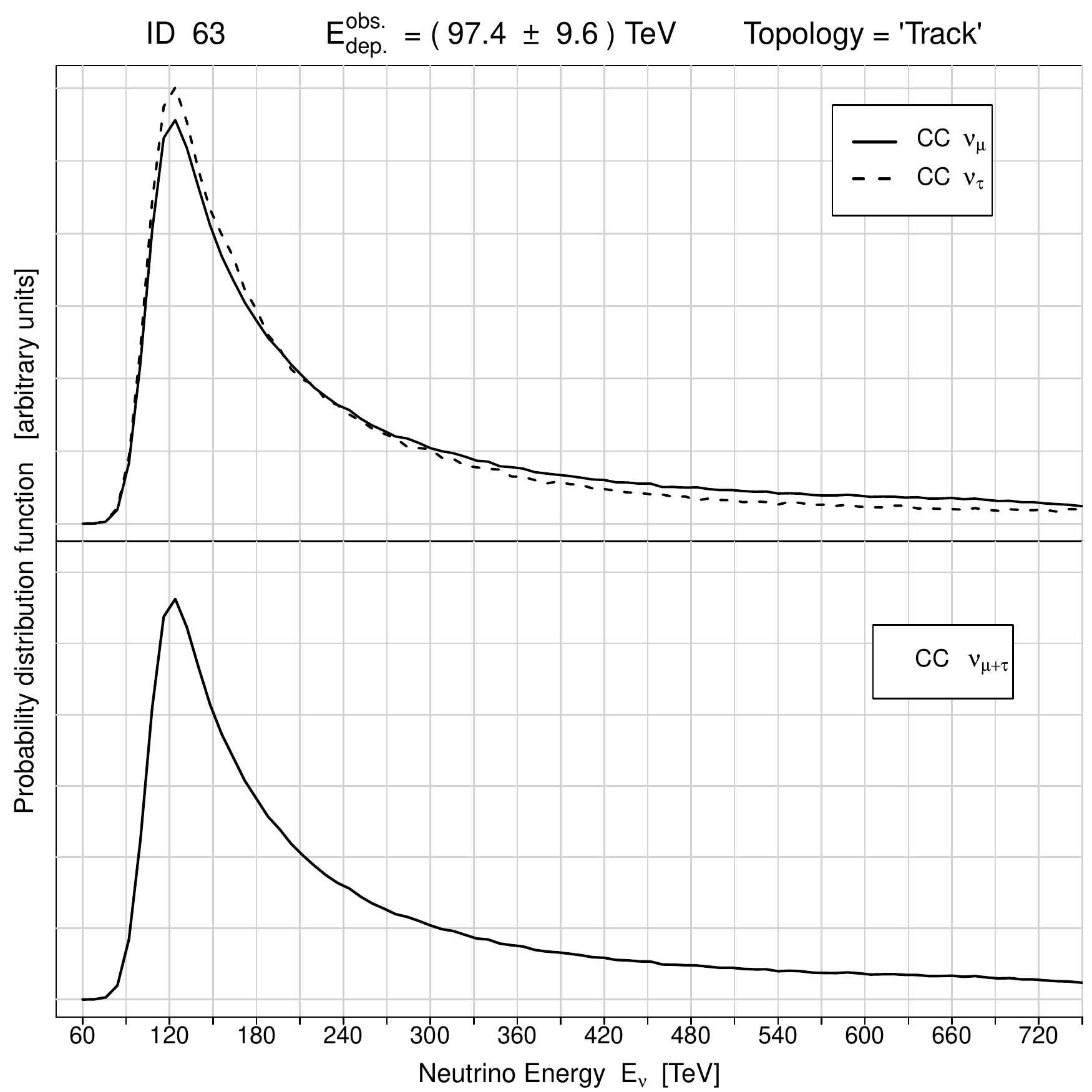}
    \includegraphics[width=0.48\linewidth,height=0.35\textheight]{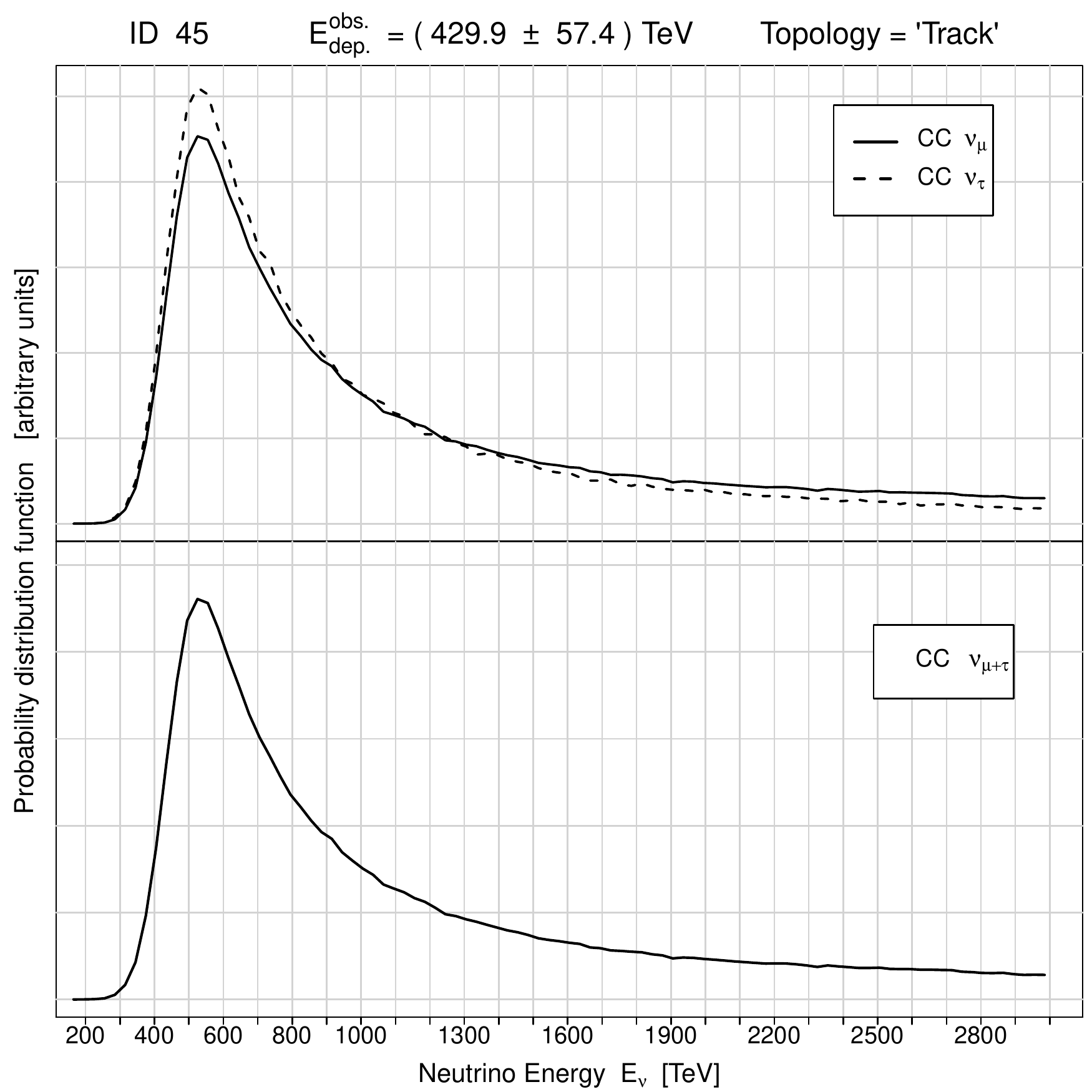} 
\caption{Posterior probability distributions $f( E_{\nu} | \ell, k, E_{dep.}^{obs.}, \text{track})$ of the neutrino energy for two track events. In the top panel the distributions assuming $\nu_{\mu}$ CC interaction (solid line) and $\nu_{\tau}$ CC interaction (dashed line) are shown. In the bottom panel the distribution $f( E_{\nu} |  E_{dep.}^{obs.}, \text{track})$, obtained marginalizing over $\ell$ and $k$,  is shown.}
\label{fig:plots_PDF_track}
\end{figure*}

\begin{figure*}
    \includegraphics[width=0.49\linewidth,height=0.3\textheight]{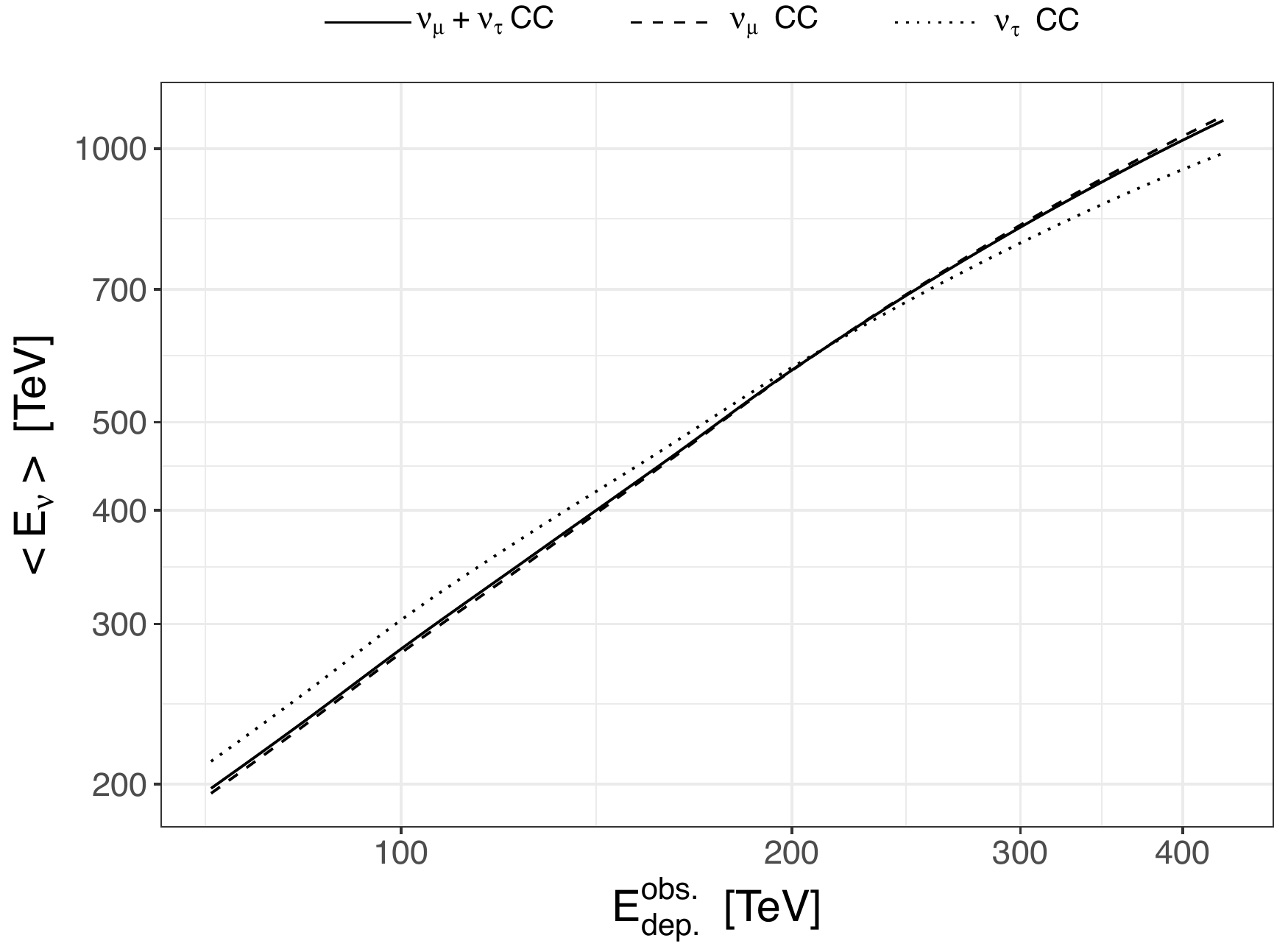}
    \includegraphics[width=0.49\linewidth,height=0.3\textheight]{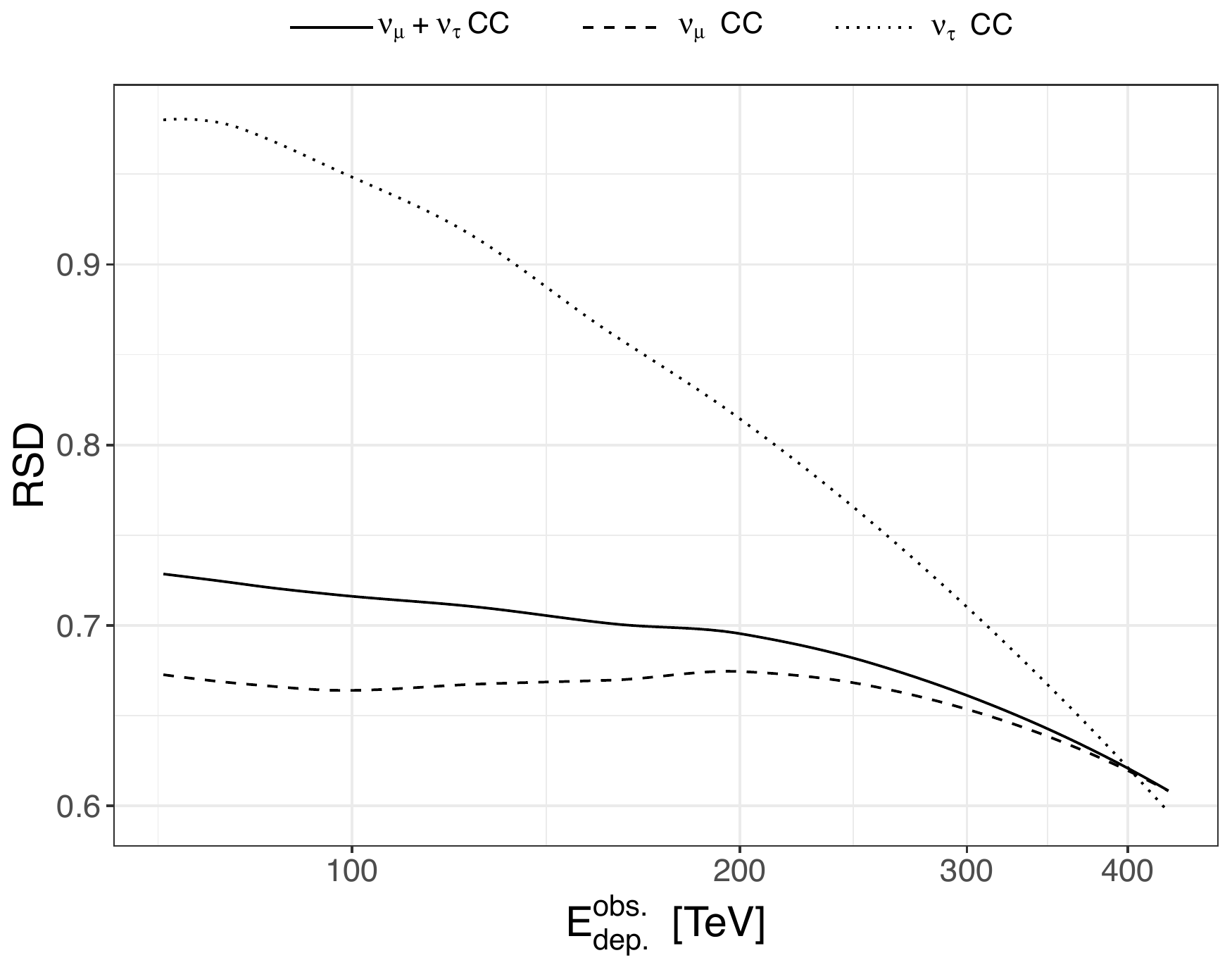}
\caption{The mean value of the neutrino energy $E_{\nu}$ (left) and its RSD value (right) are shown as a function of the observed-deposited energy in track events making no assumption (solid line), assuming $\nu_{\mu}$ CC interaction (dashed line) and $\nu_{\tau}$ CC interaction (dotted line).}
\label{fig:plots_E_track}
\end{figure*}

We show the same plots for track events: in Fig. \ref{fig:plots_PDF_track} one can find the posterior distributions for two track events, while in Fig. \ref{fig:plots_E_track} we show the neutrino-energy mean value and RSD as a function of the observed-deposited energy. 

The main difference between the energetic distribution for showers $f(E_{\nu}  | E_{dep.}^{obs.}, \text{shower})$ and tracks $f(E_{\nu}  | E_{dep.}^{obs.}, \text{track})$ is that for the latter the distribution mode is higher than the observed-deposited energy while the distribution tail is more pronounced at higher energy. This is mainly due to the fact that tracks are produced by muons, whose energy loss in the detector is only a fraction of the neutrino energy $E_{\nu}$. 

An important feature that emerges from this analysis, in particular from the right plots of Fig. \ref{fig:plots_E_shower} and \ref{fig:plots_E_track}, is that, as we approach higher observed-deposited energy, the neutrino-energy distributions become less dispersed, a fact which is illustrated by the decreasing values taken by the RSD at higher energy. This behaviour can be understood taking into account the very steeply falling of the neutrino spectrum: at higher energy the right tail of the neutrino-energy distribution becomes less pronounced, because higher energies become less frequent, and this results in a less relative dispersion in the density distribution.
The very steeply falling of neutrino spectrum (with a spectral index of $\sim 2.13$) plays also a crucial role when estimating the posterior flavor probabilities $f( \ell  | E_{dep.}^{obs.}, \text{shower})$. For instance, considering that a $\nu_{\mu}$ produces a shower only in NC interactions, one should expect \emph{a priori} that the probability for a shower event of being generated by a muonic neutrino is $\sim \, 10 \%$, being $\sim 30 \, \%$ the probability for a neutrino of scattering via NC interaction (see Fig. \ref{fig:fit_NC}) and $1/3$ the probability of being muonic. Instead our Bayesian inference gives us a value of $\sim 4 \, \% $. As mentioned above, this value, which is smaller than the expected one, can be explained only considering the neutrino spectrum and the existence of other mechanisms with better efficiency in producing a deposited energy (such as the $\nu_{e}$ CC interaction): higher energies are less frequent, thus the more particles can escape the detector after a neutrino-nucleon interaction, the less chance there is of this interaction having occurred in the detector. From this considerations it is not surprising that a neutrino producing a shower event has the best chance of being electronic: from Table \ref{tab:results} we have about $\sim 62 \, \%$ of chance that its flavor is electronic,  $\sim 4 \, \%$ muonic and $\sim 34 \, \%$ tauonic.
For track events the situation is simpler: we need to consider only $\nu_{\mu}$ and $\nu_{\tau}$ (followed by $\tau \rightarrow \nu_{\tau} \mu \nu_{\mu}$)  CC interactions. Both interactions have similar efficiency in producing a deposited energy (as illustrated by the top panels of Fig. \ref{fig:plots_PDF_track}), thus for track events when estimating the chance of being muonic or tauonic the most important thing to consider is the branching fraction for the muonic $\tau$-decay channel (see Eq. \ref{eq:br_fraction}). Our Bayesian inference for track events (see Table \ref{tab:results_track}) ends up giving to neutrinos a $\sim 87 \, \%$ chance of being muonic and $\sim 13 \, \%$ tauonic. 

Performing an inference analysis of neutrino fluxes over the whole data sample goes beyond the scope of this work, as we are interested only in inferring properties of each single neutrino event. But it is worth noticing that, combining our flavor probabilities with the observed track-to-shower ratio, we obtain that the expected flavor ratio  $(1 : 1 : 1 )_{\oplus}$ is currently disfavored. The current track-to-shower ratio for events above 60 TeV is 13/50, thus the flavor flux is approximately given by
\begin{widetext}
\begin{equation}
\sim \frac{1}{50} \left( 37 \cdot 62 \, \%  \, : \, 37 \cdot 4 \, \%  + 13 \cdot 87 \, \% \, : \, 37 \cdot 34 \, \%  + 13 \cdot 13 \, \% \right)_{\oplus} \propto (1 : 0.54 : 0.63 )_{\oplus}.
\end{equation}
\end{widetext}
This finding is in agreement with previous results obtained independently in other analyses \cite{Aartsen:2015ivb,PALOMARESRUIZ2016433, Palladino:2015zua}.
Thus the current observed track-to-shower ratio implies that the electronic flavor is almost two times more frequent than the muonic or tauonic flavor. Although it is not statistically significant at present and a complete discussion of its implications goes beyond the scope of this paper,  this result may be explained either by a misidentification of tracks as showers (according to IceCube the fraction of track misidentification is about $\sim 30 \, \%$ \cite{Aartsen:2015ivb}, while the reverse, i.e., a shower being misclassified as a track, is very rare \cite{Aartsen:2014muf}) or, even more compellingly, by some new physics that goes beyond the standard model. Therefore, a further investigation in this direction will be crucial when more data will be collected.

In this work we performed, for the first time, a detailed Bayesian inference analysis for each of the 50 high-energy neutrino events above 60 TeV detected by IceCube in 6 years of data taking. We have shown how from the observed-deposited energy and the topology event one can obtain an estimate of the neutrino energy and flavor. We have also explained how this analysis depends on the assumptions made for neutrino fluxes and for the physics involved in all processes producing shower and track events in the detector. From these assumptions we selected those prior probability distributions which seem, at present, the most reasonable ones. Further investigations in high-energy neutrino physics may change the current situation, improving our knowledge of the prior probability distribution for the parameters involved in this inference analysis. 

Neutrino astronomy has just started with IceCube providing the first evidence of astrophysical high-energy neutrinos. Inference analyses, as the one here exposed, for the properties of each high-energy neutrino have become impelling in searches for new physics and in order to shed some light on many of the questions raised by the observation of these events.  
\begingroup
 \squeezetable
\def\arraystretch{1.27}
\begin{table*}[!h]
%\rowcolors{1}{}{lightgray}
    \caption{Here we show the relevant properties of the 37 shower events with observed-deposited energy above 60 TeV. The first three columns are respectively the ID number, the observed-deposited energy and its uncertainty for each shower event. From the fourth to the seventh columns we have the mean value (mean) and the standard deviation (s.d.) of the posterior distributions of $E_{\nu}$ assuming different neutrino flavor and kind of interaction. In the last columns the probabilities for each shower event of being generated by a electronic, muonic and tauonic neutrino and of having scattered with nucleon via CC or NC interaction are shown.  }
     \begin{ruledtabular}
    \begin{tabular}{cccccccccccccccc}
        \multirow{4}{*}{ID } & \multirow{4}{*}{$E_{dep.}^{obs.}$ [TeV]} & \multirow{4}{*}{$\sigma_{E_{dep.}}$ [TeV]} & \multicolumn{8}{c}{$E_{\nu}$ [TeV]} & \multicolumn{5}{c}{prob. [$ \% $]} \\ \cline{4-11}
          & & &  \multicolumn{2}{c}{CC + NC } &  \multicolumn{2}{c}{CC} &   \multicolumn{2}{c}{CC} &  \multicolumn{2}{c}{NC} & \multirow{3}{*}{$\nu_e$ } & \multirow{3}{*}{$\nu_{\mu} $ } & \multirow{3}{*}{$\nu_{\tau} $ } & \multirow{3}{*}{CC } & \multirow{3}{*}{NC } \\ 
          & & &  \multicolumn{2}{c}{$\nu_e + \nu_{\mu} + \nu_{\tau}$}  &  \multicolumn{2}{c}{$\nu_e$} &   \multicolumn{2}{c}{$\nu_{\tau}$} &  \multicolumn{2}{c}{$\nu_e + \nu_{\mu} + \nu_{\tau}$} & & & & \\ \cline{4-11}
          & & & mean & s.d.  &  mean & s.d. &   mean & s.d. &  mean & s.d. & & & & \\
         %\hline \multicolumn{12}{c}{Shower events} \\
          \hline
2  &  117  &  15.4  &  162  &  158  &  116  &  17  &  174  &  94  &  343  &  364  &  62.2  &  4.2  &  33.5  &  87.3  &  12.7 \\
4  &  165.4  &  19.8  &  222  &  178  &  165  &  21  &  245  &  121  &  444  &  399  &  62.5  &  4.0  &  33.5  &  88.0  &  12.0 \\
9  &  63.2  &  8.9  &  92  &  112  &  64  &  8  &  94  &  48  &  201  &  271  &  60.9  &  4.5  &  34.6  &  86.5  &  13.5 \\
10  &  97.2  &  12.4  &  134  &  130  &  97  &  13  &  145  &  75  &  276  &  306  &  62.2  &  4.2  &  33.5  &  87.3  &  12.7\\
11  &  88.4  &  12.5  &  121  &  124  &  88  &  13  &  130  &  68  &  257  &  296  &  62.5  &  4.1  &  33.4  &  87.5  &  12.5\\
12  &  104.1  &  13.2  &  143  &  146  &  104  &  14  &  154  &  88  &  304  &  345  &  62.5  &  4.1  &  33.4  &  87.7  &  12.3 \\
14  &  1040.7  &  144.4  &  1195  &  379  &  1020  &  153  &  1402  &  411  &  1671  &  529  &  63.9  &  3.1  &  33.1  &  90.7  &  9.3 \\
17  &  199.7  &  27.2  &  266  &  203  &  197  &  29  &  294  &  146  &  519  &  435  &  62.2  &  4.2  &  33.6  &  87.5  &  12.5 \\
19  &  71.5  &  7.2  &  103  &  119  &  73  &  8  &  108  &  54  &  229  &  287  &  62.4  &  4.3  &  33.3  &  87.1  &  12.9\\
20  &  1140.8  &  142.8  &  1307  &  377  &  1128  &  150  &  1531  &  409  &  1789  &  512  &  64.2  &  3.0  &  32.8  &  91.1  &  8.9 \\
22  &  219.5  &  24.4  &  297  &  218  &  220  &  26  &  329  &  157  &  572  &  461  &  61.8  &  4.2  &  34.0  &  87.5  &  12.5 \\
26  &  210  &  29  &  279  &  213  &  207  &  31  &  308  &  154  &  546  &  455  &  62.2  &  4.2  &  33.6  &  87.6  &  12.4 \\
27  &  60.2  &  5.6  &  88  &  107  &  62  &  6  &  91  &  48  &  195  &  264  &  61.9  &  4.2  &  33.8  &  87.2  &  12.8 \\
30  &  128.7  &  13.8  &  176  &  152  &  130  &  15  &  192  &  98  &  354  &  350  &  62.4  &  4.1  &  33.5  &  87.8  &  12.2 \\
33  &  384.7  &  48.6  &  491  &  278  &  381  &  52  &  561  &  244  &  864  &  532  &  62.5  &  3.9  &  33.6  &  88.2  &  11.8 \\
35  &  2003.7  &  261.5  &  2085  &  346  &  1969  &  274  &  2333  &  348  &  2414  &  349  &  71.0  &  1.7  &  27.2  &  94.8  &  5.2 \\
39  &  101.3  &  13.3  &  141  &  148  &  101  &  14  &  150  &  77  &  305  &  349  &  62.1  &  4.2  &  33.6  &  87.2  &  12.8 \\
40  &  157.3  &  16.7  &  217  &  183  &  158  &  18  &  236  &  113  &  440  &  407  &  62.2  &  4.2  &  33.6  &  87.3  &  12.7 \\
41  &  87.6  &  10  &  124  &  128  &  88  &  11  &  133  &  78  &  264  &  297  &  62.3  &  4.3  &  33.4  &  87.1  &  12.9 \\
42  &  76.3  &  11.6  &  106  &  117  &  76  &  12  &  112  &  63  &  228  &  277  &  61.9  &  4.3  &  33.7  &  87.0  &  13.0 \\
46  &  158  &  16.6  &  218  &  183  &  159  &  18  &  236  &  114  &  442  &  406  &  62.0  &  4.3  &  33.7  &  87.2  &  12.8 \\
48  &  104.7  &  13.5  &  145  &  145  &  104  &  15  &  156  &  83  &  307  &  339  &  62.0  &  4.3  &  33.7  &  87.3  &  12.7 \\
51  &  66.2  &  6.7  &  95  &  105  &  67  &  7  &  102  &  61  &  204  &  244  &  61.9  &  4.4  &  33.7  &  86.9  &  13.1 \\
52  &  158.1  &  18.4  &  214  &  175  &  158  &  20  &  235  &  117  &  427  &  391  &  62.5  &  4.1  &  33.4  &  87.7  &  12.3 \\
56  &  104.2  &  10  &  145  &  133  &  106  &  11  &  157  &  78  &  298  &  311  &  62.3  &  4.1  &  33.6  &  87.6  &  12.4 \\
57  &  132.1  &  18.1  &  182  &  172  &  131  &  19  &  194  &  101  &  386  &  393  &  62.3  &  4.2  &  33.4  &  87.2  &  12.8 \\
59  &  124.6  &  11.7  &  174  &  156  &  126  &  13  &  189  &  99  &  355  &  354  &  62.2  &  4.2  &  33.6  &  87.4  &  12.6 \\
60  &  93  &  12.9  &  129  &  132  &  92  &  14  &  138  &  75  &  274  &  305  &  62.0  &  4.4  &  33.6  &  86.9  &  13.1\\
64  &  70.8  &  8.1  &  102  &  113  &  72  &  9  &  107  &  58  &  219  &  265  &  61.9  &  4.5  &  33.7  &  86.7  &  13.3\\
66  &  84.2  &  10.7  &  116  &  115  &  84  &  11  &  127  &  71  &  240  &  273  &  62.0  &  4.1  &  33.9  &  87.6  &  12.4\\
67  &  165.7  &  16.5  &  230  &  193  &  167  &  18  &  249  &  120  &  464  &  424  &  61.7  &  4.3  &  34.0  &  87.0  &  13.0\\
70  &  98.8  &  12  &  137  &  138  &  99  &  13  &  147  &  83  &  287  &  325  &  62.4  &  4.2  &  33.4  &  87.5  &  12.5\\
74  &  71.3  &  9.1  &  100  &  110  &  72  &  9  &  106  &  57  &  214  &  262  &  62.0  &  4.3  &  33.7  &  87.0  &  13.0\\
75  &  164  &  21.4  &  222  &  187  &  163  &  23  &  242  &  126  &  450  &  415  &  62.4  &  4.1  &  33.5  &  87.6  &  12.4\\
79  &  158.2  &  20.3  &  215  &  182  &  157  &  22  &  233  &  115  &  435  &  401  &  62.1  &  4.3  &  33.6  &  87.2  &  12.8\\
80  &  85.6  &  11.1  &  119  &  127  &  85  &  12  &  127  &  65  &  256  &  304  &  62.4  &  4.2  &  33.4  &  87.3  &  12.7\\
81  &  151.8  &  21.6  &  203  &  174  &  150  &  23  &  222  &  112  &  410  &  392  &  62.4  &  4.1  &  33.5  &  87.6  &  12.4
\end{tabular}
\end{ruledtabular}
\label{tab:results}
\end{table*}
\endgroup
\begingroup
 \squeezetable
  \def\arraystretch{1.27}
\begin{table*}[!h]
%\rowcolors{1}{}{lightgray}
    \caption{Same as for Table \ref{tab:results}, but for the 13 track events with observed-deposited energy above 60 TeV.}
    \begin{ruledtabular}
    \begin{tabular}{ccccccccccc}
        \multirow{4}{*}{ID } & \multirow{4}{*}{$E_{dep.}^{obs.}$ [TeV]} & \multirow{4}{*}{$\sigma_{E_{dep.}}$ [TeV]} & \multicolumn{6}{c}{$E_{\nu}$ [TeV]} & \multicolumn{2}{c}{prob. [$ \% $]} \\ \cline{4-9}
          & & & \multicolumn{2}{c}{CC}   & \multicolumn{2}{c}{CC} &  \multicolumn{2}{c}{CC}   & \multirow{3}{*}{$\nu_{\mu} $ } & \multirow{3}{*}{$\nu_{\tau} $ } \\ 
          & & & \multicolumn{2}{c}{ $ \nu_{\mu} + \nu_{\tau}$}  & \multicolumn{2}{c}{$\nu_{\mu}$ } & \multicolumn{2}{c}{ $\nu_{\tau}$} &  & \\  \cline{4-9}
          & & & mean & s.d.  & mean & s.d. &  mean & s.d.&  & \\
         %\hline \multicolumn{12}{c}{Shower events} \\
          \hline  
3	 & 78.7	 & 10.8	 & 218  &  158	 & 215  &  146	 & 232  &  221	 & 86.8	 & 13.2\\
5	 & 71.4	 & 9	 & 197  &  142	 & 195  &  131	 & 209  &  198	 & 86.8	 & 13.2\\
13	 & 252.7	 & 25.9	 & 719  &  495	 & 721  &  488	 & 703  &  538	 & 87.1	 & 12.9\\
23	 & 82.2	 & 8.6	 & 226  &  163	 & 224  &  150	 & 240  &  232	 & 86.8	 & 13.2\\
38	 & 200.5	 & 16.4	 & 571  &  395	 & 570  &  382	 & 577  &  469	 & 87.0	 & 13.0\\
44	 & 84.6	 & 7.9	 & 235  &  170	 & 233  &  154	 & 254  &  251	 & 86.7	 & 13.3\\
45	 & 429.9	 & 57.4	 & 1071  &  650	 & 1083  &  657	 & 986  &  587	 & 87.3	 & 12.7\\
47	 & 74.3	 & 8.3	 & 207  &  150	 & 205  &  136	 & 222  &  217	 & 86.8	 & 13.2\\
62	 & 75.8	 & 7.1	 & 217  &  156	 & 214  &  142	 & 235  &  229	 & 86.7	 & 13.3\\
63	 & 97.4	 & 9.6	 & 275  &  197	 & 272  &  181	 & 296  &  280	 & 86.8	 & 13.2\\
71	 & 73.5	 & 10.5	 & 200  &  149	 & 197  &  134	 & 216  &  225	 & 86.7	 & 13.3\\
76	 & 126.3	 & 12.7	 & 356  &  253	 & 352  &  235	 & 379  &  348	 & 86.7	 & 13.3\\
82	 & 159.3	 & 15.5	 & 451  &  316	 & 450  &  302	 & 463  &  395	 & 87.0	 & 13.0\\    
\end{tabular}
\end{ruledtabular}
\label{tab:results_track}
\end{table*}  
\endgroup  
\section*{Acknowledgements}

We are grateful to Giovanni Amelino-Camelia for his encouragement in writing this paper and we are grateful to Gennaro Miele for some valuable comments on an earlier version of this manuscript.
%
%
%
%
%
%
%
%
%
%
%
%
%
%\bibliographystyle{apsrev}
%\bibliography{bib/osc}

%\bibliographystyle{apsrev}
%\bibliography{my_bibliography}

\end{document}